\documentclass[prd, amsfonts, onecolumn, nofootinbib, showpacs]{revtex4}
\usepackage{graphicx, epsfig}
\usepackage{color}
\usepackage{amsmath}
\usepackage{amssymb}
\newcommand{\be}{\begin{equation}}
\newcommand{\ee}{\end{equation}}
\newcommand{\bea}{\begin{eqnarray}}
\newcommand{\eea}{\end{eqnarray}}

\newcommand{\gapp}{\mathrel{\raise.3ex\hbox{$>$}\mkern-14mu
\lower0.6ex\hbox{$\sim$}}}
\newcommand{\lapp}{\mathrel{\raise.3ex\hbox{$<$}\mkern-14mu
\lower0.6ex\hbox{$\sim$}}}
\def\bbox{{\,\lower0.9pt\vbox{\hrule \hbox{\vrule height 0.2 cm
\hskip 0.2 cm \vrule  height 0.2 cm}\hrule}\,}}

\begin{document}
\title{A note on a covariant version of Verlinde's emergent gravity}
\author{De-Chang Dai$^1$, Dejan Stojkovic$^2$}
\affiliation{$^1$ Institute of Natural Sciences, Shanghai Key Lab for Particle Physics and Cosmology, and Center for Astrophysics and Astronomy, Department of Physics and Astronomy, Shanghai Jiao Tong University, Shanghai 200240, China}
\affiliation{ $^2$ HEPCOS, Department of Physics, SUNY at Buffalo, Buffalo, NY 14260-1500}
 %%%%%%%%%%%%%%%%%%%%%%%%%%%%%%%%%%%%%%%%%%%%%%%%%%%%%%%

\begin{abstract}
\widetext
Following recent Verlinde's heuristic construction or emergent gravity, Hossenfelder wrote down the Lagrangian capturing some aspects of this theory. We point out that there is an error in calculations whose consequence is that a cosmological de Sitter space solution is not recovered from this Lagrangian. We correct this error in order to obtain the desired solution. We also show that small perturbations around de Sitter space grow, which implies that this state is not stable. However, the presence of matter and radiation might in principle provide stability.
\end{abstract}

%%%%%%%%%%%%%%%%%%%%%%%%%%%%%%%%%%%%%%%%%%%%%%%%%%

\maketitle
The idea of emergent gravity, put forward by Verlinde in \cite{Verlinde:2016toy} (see also \cite{Dai:2017qkz}), is an interesting attempt to reformulate gravity as a force of entopic origin. The main problem with this proposal is that represents just a collection of ideas without an explicit realisation. Then, in \cite{Hossenfelder:2017eoh}  a covariant version of Verlinde's gravity was formulated. For our convenience, we follow the reference \cite{Hossenfelder:2017eoh} in writing down the Lagrangian which is supposed to describe a covariant version of Verlinde's emergent gravity. We first define the quantities that enter the Lagrangian.

The central object is a vector field $u_\mu$ (also called the displacement field), with the associated elastic strain tensor
\begin{equation}
\epsilon_{\mu\nu}=\triangledown_\mu u_{\nu}+\triangledown_\nu u_\mu  ,
\end{equation}
where $\triangledown$ is the covariant derivative in the given curved background.

We can also define a dimensionless scalar
\begin{equation}
\phi=\sqrt{-u^\nu u_\nu}{L}
\end{equation}
where $\Lambda=L^{-2}$ is the cosmological constant.
Some useful shorthands are
\begin{equation}
u=\sqrt{-u^\nu u_\nu} \text{, }\epsilon=\epsilon^\nu_\nu \text{, }n^\nu= \frac{u^{\nu}}{u} .
\end{equation}

The general kinetic term for the vector field is
\begin{equation}
\chi =\alpha \triangledown_\nu u^\nu \triangledown_\kappa u^\kappa +\beta \triangledown_\nu u_\kappa \triangledown^\nu u^\kappa +\gamma  \triangledown_\nu u_\kappa \triangledown^\kappa u^\nu .
\end{equation}
In \cite{Hossenfelder:2017eoh}, the choice $\alpha =4/3$, $\beta = \gamma =-1/2$ is taken which leads to
\begin{equation}
\chi =-\frac{1}{4}\epsilon _{\mu\nu}\epsilon^{\mu\nu}+\frac{1}{3}\epsilon^2 .
\end{equation}

The total Lagrangian of the theory proposed in \cite{Hossenfelder:2017eoh} is
\begin{eqnarray} \label{L}
&&\mathcal{L}_{tot}=m_p^2 \mathcal{R}+\mathcal{L}_M+\mathcal{L}_{int}+\mathcal{L}_s \\
&&\mathcal{L}_{int}=-\frac{1}{L}u^{\mu}n^{\nu}T_{\mu\nu}=\frac{-u^{\mu}u^{\nu}}{Lu}T_{\mu\nu}\\
&&\mathcal{L}_s =\frac{m_p^2}{L^2}\chi^{3/2}-\frac{\lambda^2m_p^2}{L^4}u_\kappa u^\kappa ,
\end{eqnarray}
where $\mathcal{R}$ is the curvature of the background space, $\mathcal{L}_M$ is the Lagrangian of the normal matter, and $\mathcal{L}_{int}$ is the interaction Lagrangian.

In the way this Lagrangian is written, it cannot give a correct de Sitter space solution. There is a calculational error in Eq.~(22) of the original paper \cite{Hossenfelder:2017eoh}. One can see this when the definition for the energy momentum tensor in Eq.~(7) is applied to the Lagrangian in equation Eq.~(6) in \cite{Hossenfelder:2017eoh}. In the first parentheses on the rhs of Eq.~(22), the first two terms have an extra factor of $2$ which should not be there. Since this is a relative rather than overall sign error, the whole result in affected in a non-trivial way.

 To get the correct de Sitter space solution, one needs to modify the $\mathcal{L}_s$ part of the Lagrangian as
\begin{equation}
\mathcal{L}_s = \frac{m_p^2}{L^2}\chi^{3/2} + \frac{\lambda^2m_p^2}{L^4}(u_\kappa u^\kappa)^2 .
\end{equation}
With this modification, one can now proceed with calculations.
By definition, the energy momentum tensor for normal matter $\left(T_M\right)_{\mu \nu}$ is
\begin{equation}
\frac{\delta \mathcal{L}_M}{\delta g^{\mu\nu}}=-\frac{1}{2}\left[\left(T_M\right)_{\mu \nu}-g_{\mu\nu}\mathcal{L}_M\right] .
\end{equation}

The energy momentum tensor of the full system is
\begin{eqnarray}
&&T_{\mu\nu}=\left({T_s}\right)_{\mu\nu}+\left(T_M\right)_{\mu\nu}+\left(T_{int}\right)_{\mu\nu}\\
&&\left({T_s}\right)_{\mu\nu}=\frac{m_p^2}{L^2}\sqrt{\chi}(\frac{3}{2}\epsilon_{\mu\alpha} \epsilon^\alpha_\nu-2\epsilon_{\mu\nu} \epsilon +\chi g_{\mu\nu})-\frac{\lambda^2m_p^2}{L^4}(4u_\mu u_\nu u^\kappa u_\kappa-g_{\mu\nu} (u^\kappa u_\kappa)^2)\\
&&\left(T_{int}\right)_{\mu\nu} =\frac{4u_\mu u^\gamma \left(T_M\right)_{\nu \gamma}}{L u}+\frac{u_\nu u_\mu u^{\kappa} u^\gamma \left(T_M\right)_{\kappa \gamma}}{L u^3}-g_{\mu \nu}\frac{u^\kappa u^\gamma \left(T_M\right)_{\kappa \gamma}}{L u}
\end{eqnarray}
%Note that these expressions differ from the corresponding expressions in \cite{Hossenfelder:2017eoh} due to the calculational error in \cite{Hossenfelder:2017eoh}.

We concentrate here on the cosmological solution which is the most important if we want to reproduce the universe that we live in. We adopt the FRW form of the metric
\begin{equation}
ds^2=-dt^2 +e^{2v(t)}\left(dr^2+r^2(d\theta^2+\sin^2\theta d\phi^2)\right) ,
\end{equation}
with an ansatz for the solution
\begin{equation}
u_t=NLe^{2\xi(t)}\text{, } u_r=u_\theta=u_\phi =0 .
\end{equation}
The kinetic term becomes
\begin{equation}
\chi=N^2L^2e^{4\xi}\Big(\frac{4}{3}\dot{\xi}^2+16\dot{\xi}\dot{v}+9\dot{v}^2\Big) ,
\end{equation}
and
\begin{eqnarray}
&&\epsilon_t^t=-4NLe^{2\xi}\dot{\xi}\\
&&\epsilon^r_r=\epsilon_\theta^\theta=\epsilon_\phi^\phi=-2NLe^{2\xi}\dot{v} .
\end{eqnarray}

We can now calculate explicitly the elements of the  energy momentum tensor
\begin{eqnarray}
&&\left({T_s}\right)_t^t=\frac{m_p^2}{3}N^2e^{4\xi}\sqrt{\chi}(-20\dot{\xi}^2-96\dot{\xi}\dot{v}+27\dot{v}^2)-3N^4\lambda^2 m_p^2 e^{8\xi}\\
&&\left({T_s}\right)^r_r=\left({T_s}\right)^\theta_\theta=\left({T_s}\right)^\phi_\phi=\frac{ m_p^2}{3} N^2e^{4\xi}\sqrt{\chi}(4\dot{\xi}^2-27\dot{v}^2)+N^4\lambda^2 m_p^2 e^{8\xi}\\
\label{int-tensor}
&&\left(T_{int}\right)^t_t =2Ne^{2\xi} \left(T_M\right)_{tt}\\
&&\left(T_{int}\right)^r_r =-Ne^{2\xi} \left(T_M\right)_{tt}
\end{eqnarray}

The Einstein's equations can be written as
\begin{eqnarray}\label{ee}
&&-m_p^2 3\dot{v}^2=G^0_0\\
&&-m_p^2(2\ddot{v}+ 3\dot{v}^2)=G^i_i\\
\end{eqnarray}

In the limit when $t\rightarrow \infty$, our universe should be described as a pure de Sitter space, which fixes the geometric parameters
\begin{eqnarray} \label{dss}
&&v=\frac{t}{L^*}\\
&&\xi=0 \nonumber \\
&&N=\left(\frac{2}{9}\frac{L^*}{L}\right)^{1/3}\nonumber \\
&&\lambda^2=\frac{3^{11/3}}{2^{4/3}L^2}\left(\frac{L}{L^*}\right)^{10/3} \nonumber .
\end{eqnarray}

We now derive a self-consistency condition on the parameter $\xi$ which was not mentioned in \cite{Hossenfelder:2017eoh}.
For matter and radiation dominated universe, the normal matter energy density components of the energy momentum tensor must dominate, i.e.
\begin{equation}
|\left(T_M\right)^t{}_t|\gg |\left({T_s}\right)^t{}_t|
\end{equation}
and
\begin{equation}
|\left(T_M\right)^t{}_t|\gg |\left(T_{int}\right)^t{}_t| .
\end{equation}
We see from Eq.~(\ref{int-tensor}) that this condition can be satisfied if and only if $Ne^{2\xi}\ll 1$. If $L\approx L^*$ then $\xi$ must be negative in the radiation and matter dominated universe. This is an important condition which should not be omitted from discussion.

When dealing with new theories and Lagrangians, it is of utmost importance to check the stability of the desired solution. We showed that one can indeed correct the calculational error made by the author of \cite{Hossenfelder:2017eoh}  and modify the original Lagrangian to obtain de Sitter at late times, however, it is easy to show that this late time solution is unstable. To demonstrate this, we take small perturbation around the solution in Eq.~(\ref{dss})
\begin{eqnarray}
&&\dot{v}=\frac{1}{L^*}+\delta \dot{v}\\
&&\xi= \delta \xi  ,
\end{eqnarray}
where $\delta \dot{v}$ and $\delta \xi$ are small perturbations. The energy momentum tensor of normal matter at late time is $\left(T_M\right)^\alpha{}_\beta =0$.  We substitute this into Einstein equations in Eq.~(\ref{ee}), and keep only the first order terms. We get
\begin{eqnarray}
&&-4L^*\dot{\xi}+6L^*\delta \dot v -9\xi=0\\
&&L^*{}^2\delta \ddot{v}-6L^*\delta \dot{v}-\frac{8}{3}L^*\dot \xi -6\xi=0
\end{eqnarray}
We can eliminate the variable $\xi$ from this system to obtain
\begin{eqnarray}
\delta \ddot{v}=\frac{10}{L^*}\delta \dot{v} .
\end{eqnarray}

The solution is
\begin{equation}
\delta \dot v \sim \exp\left(\frac{10t}{L^*}\right) .
\end{equation}
This is obviously a growing mode of the perturbation. Thus, small perturbations around de Sitter space grow, which implies that this late time solution is not stable.
Note that this still does not invalidate the whole construct. A complete stability analysis at any finite time must include the contribution from matter and radiation and pertubations around the spatial components of $u$. The presence of mater and radiation might re-introduce stability. The problem however is that dark energy already dominates the universe, while the contribution from  matter and radiation rapidly diminishes. It is then not clear if the subdominant energy components can introduce stability.

\begin{acknowledgments}
D.C Dai was supported by the National Science Foundation of China (Grant No. 11433001 and 11447601), National Basic Research Program of China (973 Program 2015CB857001), the key laboratory grant from the Office of Science and Technology in Shanghai Municipal Government (No. 11DZ2260700) and  the Program of Shanghai Academic/Technology Research Leader under Grant No. 16XD1401600. D.S. was partially supported by the US National Science Foundation, under Grant No. PHY-1417317.
\end{acknowledgments}

\end{document}